\documentclass[12pt]{iopart}

\usepackage{enumerate}

\usepackage{amsthm}

\newcommand{\be}{\begin{equation}}
\newcommand{\ee}{\end{equation}}
\newcommand{\ba}{\begin{eqnarray}}
\newcommand{\ea}{\end{eqnarray}}
\eqnobysec

\begin{document}

\title[A classification of spherically symmetric spacetimes]
{A classification of spherically symmetric spacetimes}

\author{Brian O J Tupper$^\sharp$, Aidan J Keane$^\natural$ and Jaume Carot$^\flat$}
\address{$^\sharp$Department of Mathematics and Statistics, University of
New Brunswick\\Fredericton, New Brunswick, Canada E3B 5A3.\\
$^\natural$87 Carlton Place, Glasgow G5 9TD, Scotland, UK.
\\$^\flat$Departament de F\'{i}sica, Universitat de les Illes Balears,
Cra. Valldemossa km 7.5, E-07122 Palma de Mallorca, Spain.
}

\eads{\mailto{bt32@rogers.com}, \mailto{aidan@countingthoughts.com}, \mailto{jcarot@uib.es}}

\begin{abstract}
A complete classification of locally spherically symmetric four-dimensional Lorentzian spacetimes is given in terms of their local conformal symmetries. The general solution is given in terms of canonical metric types and the associated conformal Lie algebras. The analysis is based upon the local conformal decomposition into 2+2 reducible spacetimes and the Petrov type. A variety of physically meaningful example spacetimes are discussed.
\end{abstract}


\section{Introduction}
Locally spherically symmetric four-dimensional Lorentzian spacetimes are locally conformally related to 2+2 reducible spacetimes \cite{carot2002, tupper1996}. This provides a convenient framework for the classification of spherically symmetric spacetimes in terms of their local conformal symmetries. The Petrov type of a spherically symmetric spacetime is that of the underlying 2+2 reducible spacetime which is type $D$ or $O$.

Previous work in this area \cite{qadir1995} - \cite{maartens1995} has been based upon direct integration of the conformal Killing equations involving, in general, many functions of two independent variables: Qadir and Ziad \cite{qadir1995} classified spherically symmetric spacetimes according to their isometries; Moopanar and Maharaj \cite{moopanar2010} gave the formal solution for the conformal symmetries of spherically symmetric spacetimes and presented some applications; Maartens et al \cite{maartens1995} classified the conformal symmetries of {\it static} spherically symmetric spacetimes; Shabbir et al \cite{shabbir2009} considered the classification according to the conformal symmetries but the work contains some errors.

This work generalizes the isometry classification of \cite{qadir1995} by widening the scope to conformal symmetries. Spherically symmetric spacetimes are constructed from an underlying 2+2 reducible metric multiplied by a conformal function. Thus, by taking advantage of the work on 2+2 reducible spacetimes \cite{carot2002} and the Defrise-Carter theorem \cite{defrise1975,hall1991}, the classification problem is reduced to finding the various functional forms of a {\it single} function of two independent variables and an underlying 2+2 reducible metric. The canonical forms for the metrics, the associated generators of the conformal symmetry Lie algebras and their conformal type
(isometric, homothetic etc) are used as the basis of the construction. Further, the classification is based on types $D$ or $O$ separately from the outset in contrast to the existing schemes \cite{qadir1995} - \cite{shabbir2009}. The analysis constitutes a {\it general} solution to the classification problem and is independent of the field equations. It applies to solutions of standard General Relativity and theories such as $f(R)$ Gravity, e.g., \cite{capozziello2007, capozziello2012}.

Let $(M,g)$ denote a four-dimensional spacetime manifold with Lorentzian metric $g_{ab}$.
$\mathcal{L}_X$ denotes the Lie derivative operator with respect to a vector field $X$ on $M$.
For Lie algebras ${\cal A}$ and ${\cal B}$,  the notation ${\cal A} \supset {\cal B}$ means ${\cal B}$
is a subalgebra of ${\cal A}$.
A vector field $\xi$ is said to be a {\it conformal Killing vector field} (CKV) if and only if
\be
(\mathcal{L}_\xi g)_{ab} = 2 \psi g_{ab}
\label{eq:ckv}
\ee
where $\psi$ is some function of the coordinates ({\it conformal scalar}). When
$\psi$ is not constant the CKV is said to be {\it proper}, and if $\psi_{;ab}=0$ the
CKV is a {\it special} CKV (SCKV). When $\psi$ is a constant, $\xi$ is a {\it homothetic vector field} (HKV) and when the constant $\psi$ is non-zero $\xi$ is a {\it proper} HKV. When $\psi = 0$, $\xi$ is a {\it Killing vector field} (KV). The set of all CKV (respectively, SCKV, HKV and KV) form a finite-dimensional Lie algebra denoted by ${\cal C}$ (respectively, ${\cal S}$, ${\cal H}$ and ${\cal G}$).

The line element of a general spherically symmetric spacetime $(M,g)$ is \cite{KSMH2}
\be
ds^2 = -\e^{2 \nu(t,r)} dt^2 + \e^{2 \lambda(t,r)} dr^2 + Y^2(t,r) d\Omega^2
\ee
where $d\Omega^2 = d\theta^2 + \sin^2 \theta d\phi^2$. Rescaling the metric by $Y^2$ we have
\be
ds^2 = Y^2 d \Sigma^2
\label{eq:conformalrescaling}
\ee
where
\ba
d \Sigma^2 = d \sigma_1^2 + d \sigma_2^2
\label{eq:dsigmasquared}
\\
d \sigma_1^2 = Y^{-2} (-\e^{2 \nu} dt^2 + \e^{2 \lambda} dr^2),
\qquad
d \sigma_2^2 = d\Omega^2.
\ea
The spacetime $(M,g)$ is thus {\em 2+2 conformally reducible} \cite{carot2002} and the line element $d \Sigma^2$ corresponds to the metric of the underlying 2+2 reducible spacetime $(M, \hat{g})$.
$(M,\hat{g})$ has (locally) the product manifold structure, $M=V_1 \times V_2$; $V_1$ and $V_2$ being the factor submanifolds endowed with 2-dimensional metrics, $h_1$ and $h_2$ (associated line elements  $d\sigma_1^2 $ and $d\sigma_2^2$), $\hat g$ then being $\hat g= h_1\oplus h_2$. We shall refer to those as the \emph{2-spaces} $(V_1,h_1)$ and $(V_2,h_2)$. Thus, (\ref{eq:conformalrescaling}) can be written
\be
g = Y^2 \hat g.
\ee
 Of course, $h_1$ is necessarily of Lorentz signature. In general $d \sigma_1^2$ admits no KVs whereas $d \sigma_2^2$ admits the ${\cal G}_3$ Lie algebra $so(3)$ with basis,
\ba
\zeta_{\delta_1} = \cos\phi \partial_\theta - \cot\theta \sin\phi \partial_\phi,
\qquad
\zeta_{\delta_2} = -\sin\phi \partial_\theta -\cot\theta \cos\phi \partial_\phi
\nonumber\\
\zeta_{\delta_3} = \partial_\phi.
\label{eq:g3sphericalsymmetry}
\ea
The underlying 2+2 conformally reducible spacetime $d \Sigma^2$ is also spherically symmetric, i.e., it admits the $so(3)$ Lie algebra above, since the function $Y$ depends upon $t$ and $r$ only. In order to take advantage of the work on 2+2 reducible spacetimes carried out in \cite{carot2002}, the line element of $(M,g)$ is written in terms of null coordinates $u, v$, i.e.,
\be
ds^2 = Z^2(u,v)[2 f(u,v) du dv + d\Omega^2]
\ee
where $Z(u,v) = Y(t,r)$ and
\be
d \sigma_1^2 = 2 f(u,v) du dv,
\qquad
d \sigma_2^2 = d\Omega^2.
\ee
The spacetime $(M,g)$ is type $O$ when
\be
f(u,v) = 2 (u + v)^{-2}
\label{eq:ftypeo}
\ee
and type $D$ otherwise \cite{carot2002}. Given a CKV $\xi$ of the metric $\hat{g}$
\be
(\mathcal{L}_\xi \hat{g})_{ab} = 2 \hat \psi \hat{g}_{ab}
\ee
then, given that $Z$ is a function of $u$ and $v$ only, the general relationship between the conformal scalars is
\be
\psi = Z^{-1} (\xi^u Z_{,u} + \xi^v Z_{,v}) + \hat \psi.
\label{eq:conformalscalarrelationship}
\ee
In \cite{carot2002} it was shown that spherically symmetric spacetimes of the form $(M, \hat{g})$ cannot admit a HKV. When the spacetime $(M,g)$ is restored by multiplying $(M, \hat{g})$ by $Z^2(u,v)$ the members of the $so(3)$ basis (\ref{eq:g3sphericalsymmetry}) remain as KVs while the KVs and proper CKVs admitted by $(M, \hat{g})$ will, in general, become proper CKVs of $(M,g)$. However, depending on the metric $h_1$ and function $Z(u,v)$, one (or more) may be SCKVs, HKVs or KVs. Thus, the conformal symmetries of $(M,g)$ can be deduced from those of $(M, \hat{g})$ and the function $Z(u,v)$, providing a classification scheme for the spacetimes $(M,g)$, the classes being determined by the properties of the conformal scalars $\psi$. It is shown below that proper SCKVs arise in two specific spacetime classes only. The details of the calculations for HKVs and KVs are not given but the procedure involves solving the condition $\psi = c$ for both $c = 0$ and $c \ne 0$ for each member of the algebra whilst being aware that the corresponding pair $\{ \hat{g}, Z(u,v) \}$ may admit further symmetry either {\it automatically} or in {\it special cases}.

Coley and Tupper \cite{coley1992} showed that a 2+2 reducible spacetime $(M, \hat{g})$ cannot admit a proper CKV unless it is type $O$ in which case $d\sigma_1^2$ and $d\sigma_2^2$ each must be of constant curvature $k_1$, $k_2$, respectively, with $k_1+k_2=0$. Further, for type $O$ spacetimes dim ${\cal C} = 15$, dim ${\cal H} \le 11$ and dim ${\cal G} \le 10$, see \cite{hall1990}. In \cite{coley1989} it is shown that the only spherically symmetric spacetimes admitting proper SCKV are Minkowski spacetime and the type $D$ spacetime
\be
ds^2 = -dt^2 +dr^2 + t^2 d\Omega^2.
\label{eq:typedspacetimewithsckv}
\ee

In the case of type $D$ spacetimes, since the underlying spacetime $(M, \hat{g})$ cannot admit HKVs or proper CKVs, the spacetimes admit only KVs so that $\hat \psi = 0$. The ${\cal G}_r$ of $(M, \hat{g})$ is composed of the ${\cal G}$ of the 2-spaces with metrics $d\sigma_1^2$ and $d\sigma_2^2$, i.e., if $\zeta^a= (\zeta^0, \zeta^1)$ is a KV of $d\sigma_1^2$, then $\xi^a= (\zeta^0, \zeta^1,0,0)$ is a KV of $(M,\hat{g})$, etc. We remark further that for type $D$ spacetimes dim ${\cal G} \le$ dim ${\cal C} \le 6$ and dim ${\cal H} \le 5$ \cite{hall1990, hallsteele1990}. In the case of type $D$ spacetimes $(M, \hat{g})$ the conformal Lie algebra is the direct sum
\be
{\cal G}_{r+3}(M) = {\cal G}_r(V_1) \oplus {\cal G}_3 (V_2)
\label{eq:directsum1}
\ee
where $0 \le r \le 3$ and ${\cal G}_3 (V_2)$ is the $so(3)$ basis (\ref{eq:g3sphericalsymmetry}).
The corresponding spacetime $(M, g)$ has conformal Lie algebra
\be
{\cal C}_{r+3}(M) = {\cal C}_r(V_1) \oplus {\cal G}_3 (V_2).
\label{eq:directsum2}
\ee
The problem of finding and classifying the conformal algebras of type $D$ spherically symmetric spacetimes becomes the simple problem of determining the canonical functions $f(u,v)$ and corresponding Killing algebras ${\cal G}_r (V_1)$ of the 2-spaces $(V_1,h_1)$ along with the appropriate functions $Z(u,v)$.

For type $O$ spacetimes the underlying 2+2 reducible spacetime $(M, \hat{g})$ with $f(u,v)$ given by (\ref{eq:ftypeo}) admits a ${\cal C}_{15} \supset {\cal G}_6$. The classification scheme is similar to the type $D$ case except it is necessary to consider not only the ${\cal G}_r (V_1)$ but the entire ${\cal C}_{15} (M)$.

In this work we classify the spacetimes $(M,g)$ according to their conformal algebras ${\cal C}_r$ which is governed by the properties of the pair $\{ \hat{g}, Z(u,v) \}$. In section \ref{sec:2spaces} the relevant 2-spaces $(V_1,h_1)$, functions $Z(u,v)$ and the corresponding ${\cal C}_r$ are determined. type $D$ spacetimes are dealt with in section \ref{sec:typed} and as remarked above, it is necessary only to consider the ${\cal G}_r (V_1)$ algebras of the 2-space $(V_1,h_1)$. type $O$ spacetimes are dealt with in section \ref{sec:typeo}. A subset of the results of section \ref{sec:2spaces} are relevant in the type $O$ case.

\section{The geometry of the 2-spaces $(V_1,h_1)$}\label{sec:2spaces}
The metric $h_1$ is necessarily of Lorentz signature. The relevant metric classes and ${\cal G}_r$ for $(V_1,h_1)$ are as follows \cite{carot2002}. The function $G$ is an {\it arbitrary} function of its arguments.
\begin{enumerate}[I]
\item $h_1$ is of constant curvature $K_0$. We distinguish two subclasses:
\begin{enumerate}
\item $K_0 \ne 0$, $f(u,v) = 2 K_0^{-1} (u+v)^{-2}$ with ${\cal G}_3$ basis
    \[ \fl
    \xi_1 = u^2 \partial_u - v^2 \partial_v, \qquad \xi_2 = u \partial_u + v \partial_v, \qquad
    \xi_3 = \partial_u - \partial_v.
    \]
\item $K_0 = 0$, $f(u,v) = \epsilon$ where $\epsilon = \pm 1$, with ${\cal G}_3$ basis
    \[ \fl
    \eta_1 = \partial_u, \qquad \eta_2 = \partial_v, \qquad
    \eta_3 = u \partial_u - v \partial_v.
    \]
\end{enumerate}

\item $h_1$ is not of constant curvature.
\begin{enumerate}
    \item $f(u,v) = G(u+v)$ with ${\cal G}_1$
    \[ \fl
    \omega = \partial_u - \partial_v.
    \]
    \item $f(u,v) = G(uv)$ with ${\cal G}_1$
    \[ \fl
    \tau = u \partial_u - v \partial_v.
    \]

    \item $f(u,v)$ is none of the above. In this case $h_1$ admits no ${\cal G}_r$

    \end{enumerate}

\end{enumerate}
Class II(c) corresponds to the case where the spacetime $(M,g)$ admits only the maximal ${\cal G}_3$ (\ref{eq:g3sphericalsymmetry}). We emphasize that if $K_0 = 1$ in subclass I(a) then the spacetime $(M,g)$ is of type $O$ and that all other subclasses correspond to $(M,g)$ being of type $D$.
The absence of a ${\cal G}_2$ amongst the possibilities is explained by the fact that no 2-space can admit a maximal ${\cal G}_2$, i.e., if it admits a ${\cal G}_2$ it necessarily admits a ${\cal G}_3$ and is then of constant curvature \cite{KSMH2}. We remark that the metric subclass II(b) corresponds to a metric which admits a ${\cal G}_1$ with a fixed point. In all other classes {\it at least one of} the KVs have no fixed points so that one can choose coordinates adapted to it. Details regarding the metrics admitting CKVs with fixed points can be found in section 3 of \cite{carot2002} and only the case (4) from that section is relevant here. (All other cases correspond to either positive definite metrics or HKVs which are not relevant here.)

For metric class I and classes II(a), II(b) above, an arbitrary function $Z(u,v)$ will lead to a ${\cal C}_3$ or ${\cal C}_1$ respectively. In the following we determine the pertinent functions $Z(u,v)$ which will lead to ${\cal G}_r$ or ${\cal H}_r$ for all possible choices of CKVs, and linear combinations. (There is no need to consider ${\cal S}_r$ here since we have already established that  (\ref{eq:typedspacetimewithsckv}) is the only relevant spacetime.)

In each case the CKV $\xi$ and the conditions upon the function $Z(u,v)$ for ${\cal G}$ or ${\cal H}$ subalgebras are given. In the following $a$, $b$ and $c$ are arbitrary non-zero constants, $F$ denotes an {\it arbitrary} function of its arguments, and we define
\ba
U_\pm = (u + v)(a^2 \pm uv)^{-1}.
\label{eq:shorthandnotation1}
\ea
We emphasize that here we list only those subgroups arising from the ${\cal G}(V_1)$, i.e., we ignore the $so(3)$ subalgebra which is, of course, common to all classes.

\subsection*{Metric class I(a)}

\begin{enumerate}

\item ${\cal G}_3(V_1)$ if  $Z = constant$.

\item $\xi_3 \in {\cal G}(V_1)$ if $Z = F(u+v)$.

\item $\xi_2 \in {\cal H}(V_1)$ with $\psi = c$ and $\xi_3 \in {\cal G}(V_1)$ if $Z = (u + v)^c$.

\item $\xi_3 \in {\cal H}(V_1)$ with $\psi = c$ if $Z = \exp(-cv) F(u+v)$.

\item $\xi_1 + a^2 \xi_3 \in {\cal G}(V_1)$ if $Z = F(U_-)$.

\item $\xi_1 + a^2 \xi_3 \in {\cal H}(V_1)$ with $\psi = c$ if $Z = \exp [-ca^{-1} \arctan (v/a)] F(U_-)$.

\item $\xi_1 - a^2 \xi_3 \in {\cal G}(V_1)$ if $Z = F(U_+)$.

\item $\xi_1 - a^2 \xi_3 \in {\cal H}(V_1)$ with $\psi = c$ if $Z = [(a+v)/(a-v)]^{c/2a} F(U_+)$.

\item ${\cal C}_3(V_1)$ if $Z$ is an arbitrary function.

\end{enumerate}
Let $l$, $m$, $n$ be constants. Those subclasses corresponding to $\xi_1$, $\xi_2$, and linear combinations $\xi_1 + m \xi_2$, $\xi_2 + n \xi_3$ and $\xi_1 + m \xi_2 + n \xi_3$ are not listed because they are equivalent to one of the above via an appropriate coordinate transformation. The essence of the transformation involves recognizing the components in the general case as $\xi^u = lu^2 - 2mu-n$, $\xi^v = lv^2 + 2mv-n$ and completing the square. For the case $l \ne 0$ the quadratic terms can be replaced by terms of the form $u^2 - a^2$, $u^2 + a^2$ or $u^2$ and similar for the $v$ terms. Furthermore, $(u+v)^{-2} du dv$ is {\it invariant} under the inversion $u \mapsto u^{-1}$, $v \mapsto v^{-1}$. It follows that the only CKVs $\xi$ to consider are those with $\xi^u$ given by $1$, $u^2 - a^2$ or $u^2 + a^2$. For the case $l = 0$ the quadratic expressions become linear and analogous simplifications can be made.

\subsection*{Metric class I(b)}
\begin{enumerate}

\item ${\cal G}_3(V_1)$ if  $Z = constant$.

\item $\eta_3 \in {\cal G}(V_1)$ if $Z = F(uv)$.

\item $\eta_3 \in {\cal H}(V_1)$ with $\psi = c$ if $Z = v^{-c} F(uv)$.

\item $\eta_1 + a \eta_2 \in {\cal G}(V_1)$ if $Z = F(au - v)$.

\item $\eta_1 + a \eta_2 \in {\cal H}(V_1)$ with $\psi = c$ if $Z = \exp(cv/a) F(au - v)$.

\item ${\cal C}_3(V_1)$ if $Z$ is an arbitrary function.

\end{enumerate}
Those subclasses corresponding to $\eta_1$, $\eta_2$, $\eta_1 + a \eta_3$, $\eta_2 + a \eta_3$, and $\eta_1 + a \eta_2 + b \eta_3$ are not listed because they are equivalent to one of the above by adjustment of linear combinations or via an appropriate coordinate transformation.

\subsection*{Metric class II(a)}
\begin{enumerate}

\item ${\cal G}_1 (V_1)$ if  $Z = constant$.

\item $\omega \in {\cal G}(V_1)$ if $Z = F(u+v)$.

\item $\omega \in {\cal H}(V_1)$ with $\psi = c$ if $Z = \exp (-cv) F(u+v)$.

\item ${\cal C}_1 (V_1)$ if $Z$ is an arbitrary function.

\end{enumerate}

\subsection*{Metric class II(b)}
\begin{enumerate}

\item ${\cal G}_1 (V_1)$ if  $Z = constant$.

\item $\tau \in {\cal G}(V_1)$ if $Z = F(uv)$.

\item $\tau \in {\cal H}(V_1)$ with $\psi = c$ if $Z = v^{-c} F(uv)$.

\item ${\cal C}_1 (V_1)$ if $Z$ is an arbitrary function.

\end{enumerate}
Note that the coordinate transformation $u \mapsto \e^u$, $v \mapsto \e^v$ transforms the metric class II(b) and the KV $\tau$ into metric class II(a) and the KV $\omega$. However, the transformation is not valid at the fixed point $u = v = 0$ of $\tau$. Thus subclass II(a) is a region of subclass II(b) that does not include the fixed point.

\section{Type $D$ spacetimes}\label{sec:typed}
The complete classification for the type $D$ spacetimes $(M,g)$ based on the classes in section \ref{sec:2spaces} is given in table \ref{tab:typeD}. We have already stated that (\ref{eq:typedspacetimewithsckv}) is the only type $D$ spherically symmetric spacetime $(M,g)$ that admits a proper SCKV. Putting $t = u + v$, $r = v - u$, (\ref{eq:typedspacetimewithsckv}) is
\be
ds^2 = -4 du dv + (u + v)^2 d\Omega^2
\ee
which corresponds to metric class I(a) with $K_0 = -1$ and $Z = u+v$. This spacetime $(M,g)$ admits a ${\cal S}_6 \supset {\cal H}_5 \supset {\cal G}_4$ with basis $\{ \xi_1, \xi_2, \xi_3 \}$ and conformal scalars $\psi_1 = 2(u+v), \psi_2 = 2, \psi_3 = 0$ respectively. This has been included as class I(a)(x) in table \ref{tab:typeD}.

As we have already stated for type $D$ spacetimes, the only ${\cal C}_r$ admitted by $(M, \hat{g})$ is necessarily a ${\cal G}_r$ with $so(3)$ isometry subalgebra being given by (\ref{eq:g3sphericalsymmetry}). Note that class II(c) represents a spacetime admitting only the isometry algebra $so(3)$ and no further conformal symmetries.

It follows from (\ref{eq:directsum2}) and the results of section \ref{sec:2spaces} that the conformal Lie algebras for the type $D$ spacetimes $(M,g)$ are either 3, 4 or 6-dimensional. In the 6-dimensional case the ${\cal C}_3(V_1)$ subalgebras are
\ba
&\mbox{I(a)} \qquad [\xi_1, \xi_2] = - \xi_1, \qquad [\xi_2, \xi_3] = - \xi_3, \qquad [\xi_1, \xi_3] = - 2 \xi_2
\nonumber\\
&\mbox{I(b)} \qquad [\eta_1, \eta_2] = 0, \qquad [\eta_2, \eta_3] = - \eta_2, \qquad [\eta_1, \eta_3] =  \eta_1
\nonumber
\ea
the latter being the Euclidean algebra $E(2)$. In both cases the orbits of the ${\cal C}_3(V_1)$ are 2-dimensional. In the 4-dimensional case the orbit of ${\cal C}_1(V_1)$ is 1-dimensional, except at the fixed point in case II(b).

\begin{table}
\caption{\label{tab:typeD}The type $D$ metrics and corresponding algebras. The functions $F$ and $G$ are {\it arbitrary} functions of their arguments. The $so(3)$ isometry subalgebra (\ref{eq:g3sphericalsymmetry}) is not listed since it is common to all classes. For a spherically symmetric spacetime admitting a homothety, the homothety algebra contains an isometry subalgebra with dimension one less than the dimension of the homothety algebra, i.e., ${\cal H}_r \supset {\cal G}_{r-1}$. The notation $\xi_i \in {\cal H}$ denotes that $\xi_i$ is a proper HKV with $\psi = c \ne 0$, and $\xi_i \in {\cal S}$ denotes that $\xi_i$ is a proper SCKV. The constant $K_0 \ne 1$ and $a$ is an arbitrary non-zero constant. The quantity $U_\pm$ is defined in equation (\ref{eq:shorthandnotation1}).}
\footnotesize\rm
\begin{tabular*}{\textwidth}{@{}l*{15}{@{\extracolsep{0pt plus12pt}}l}}
\br
Class & & Algebra & & Metric $ds^2$\\
\mr

I(a) & (i) & ${\cal G}_6$ & $\xi_1, \xi_2, \xi_3 \in {\cal G}$ & $d\Sigma^2 = 4 K_0^{-1}(u+v)^{-2} du dv + d\Omega^2$ \\

& (ii) & ${\cal C}_6 \supset {\cal G}_4$ & $\xi_3 \in {\cal G}$ & $F^2(u+v) d\Sigma^2$ \\

& (iii) & ${\cal C}_6 \supset {\cal H}_5$ & $\xi_2 \in {\cal H}$, $\xi_3 \in {\cal G}$ & $(u+v)^c d\Sigma^2$ \\

& (iv) & ${\cal C}_6 \supset {\cal H}_4$ & $\xi_3 \in {\cal H}$ & $\e^{-2cv} F^2(u+v) d\Sigma^2$ \\

& (v) & ${\cal C}_6 \supset {\cal G}_4$ & $\xi_1 + a^2 \xi_3 \in {\cal G}$ & $F^2(U_-) d\Sigma^2$ \\

& (vi) & ${\cal C}_6 \supset {\cal H}_4$ & $\xi_1 + a^2 \xi_3 \in {\cal H}$ & $\exp[2ca^{-1} \arctan(v/a)] F^2(U_-) d\Sigma^2$ \\

& (vii) & ${\cal C}_6 \supset {\cal G}_4$ & $\xi_1 - a^2 \xi_3 \in {\cal G}$ & $F^2(U_+) d\Sigma^2$ \\

& (viii) & ${\cal C}_6 \supset {\cal H}_4$ & $\xi_1 - a^2 \xi_3 \in {\cal H}$ & $[(a+v)/(a-v)]^{-c/a} F^2(U_+) d\Sigma^2$ \\

& (ix) & ${\cal C}_6 \supset {\cal G}_3$ & - & $F^2(u,v) d\Sigma^2$ \\

& (x) & ${\cal S}_6 \supset {\cal H}_5$ & $\xi_1 \in {\cal S}$, $\xi_2 \in {\cal H}$, $\xi_3 \in {\cal G}$ & $(u+v)^2 d\Sigma^2$ \\

& & & \\

I(b) & (i) & ${\cal G}_6$ & $\eta_1, \eta_2, \eta_3 \in {\cal G}$ & $d\Sigma^2 = 2 \epsilon du dv + d\Omega^2$ \\

& (ii) & ${\cal C}_6 \supset {\cal G}_4$ & $\eta_3 \in {\cal G}$ & $F^2(uv) d\Sigma^2$ \\

& (iii) & ${\cal C}_6 \supset {\cal H}_4$ & $\eta_3 \in {\cal H}$ & $v^{-2c} F^2(uv) d\Sigma^2$ \\

& (iv) & ${\cal C}_6 \supset {\cal G}_4$ & $\eta_1 + a \eta_2 \in {\cal G}$ & $F^2(au-v) d\Sigma^2$ \\

& (v) & ${\cal C}_6 \supset {\cal H}_4$ & $\eta_1 + a \eta_2 \in {\cal H}$ & $e^{2cv/a} F^2(au-v) d\Sigma^2$ \\

& (vi) & ${\cal C}_6 \supset {\cal G}_3$ & - & $F^2(u,v) d\Sigma^2$ \\

& & & \\

II(a) & (i) & ${\cal G}_4$ & $\omega \in {\cal G}$ & $d\Sigma^2 = 2 G(u+v) du dv + d\Omega^2$ \\

& (ii) & ${\cal G}_4$ & $\omega \in {\cal G}$ & $F^2(u+v) d\Sigma^2$ \\

& (iii) & ${\cal H}_4$ & $\omega \in {\cal H}$ & $\e^{-2cv} F^2(u+v) d\Sigma^2$ \\

& (iv) & ${\cal C}_4 \supset {\cal G}_3$ & - & $F^2(u,v) d\Sigma^2$ \\

& & & \\

II(b) & (i) & ${\cal G}_4$ & $\tau \in {\cal G}$ & $d\Sigma^2 = 2 G(uv) du dv + d\Omega^2$ \\

& (ii) & ${\cal G}_4$ & $\tau \in {\cal G}$ & $F^2(uv) d\Sigma^2$ \\

& (iii) & ${\cal H}_4$ & $\tau \in {\cal H}$ & $v^{-2c} F^2(uv) d\Sigma^2$ \\

& (iv) & ${\cal C}_4 \supset {\cal G}_3$ & - & $F^2(u,v) d\Sigma^2$ \\

& & & \\

II(c) & & ${\cal G}_3$ & - & $F^2(u,v) [2 G(u,v) du dv + d\Omega^2]$ \\

\br
\end{tabular*}
\end{table}
The classification has been achieved in terms of null coordinates but in most cases the transformation from standard $t$, $r$ coordinates is straightforward.

\subsection{Example: Schwarzschild spacetime}
Perhaps the most well known spherically symmetric spacetime is the Schwarzschild spacetime. The maximally extended metric is given in Kruskal-Szekeres coordinates as
\be
ds^2 = -32M^3 r^{-1} \e^{-r/2M} dudv + r^2 d\Omega^2
\label{eq:kruskalszekeres}
\ee
where the function $r$ is given implicitly by
\be
uv = (1-r/2M) \e^{r/2M}.
\label{eq:kruskalszekeresuv}
\ee
This is clearly a class II(b)(ii) spacetime and so admits a maximal ${\cal G}_4$, i.e., the spacetime admits no proper HKV or proper CKVs, although this has been remarked upon before \cite{tupper1996}. The fixed point of the KV occurs at $u = v = 0$.

The metric (\ref{eq:kruskalszekeres}) can be written explicitly in terms of $u$ and $v$ by noting that the expression (\ref{eq:kruskalszekeresuv}) implies that
\[
r/2M - 1 = W(- \e^{-1} uv)
\]
where $W(x)$ is the Lambert W-function \cite{corless96}. Thus the metric takes the explicit null coordinate form
\be
ds^2 = -16 M^2 (1+W)^{-1} \e^{-(1+W)} dudv + 4 M^2 (1+W)^2 d\Omega^2
\ee
where $W = W(- \e^{-1} uv)$. The solution exists on the entire principal branch of $W(x)$ from $-1 \le W < \infty$ and is valid for all $r \ge 0$. The fixed point lies on the event horizon $r = 2M$, i.e., $uv=0$ which implies that $W = 0$.

The coordinate transformation $u \mapsto \e^u$, $v \mapsto \e^v$ transforms the metric into that of a class II(a)(ii) spacetime. Equation (\ref{eq:kruskalszekeresuv}) becomes
\be
\e^{u+v} = (1-r/2M) \e^{r/2M}
\ee
which shows that the coordinates cover only the region $r<2M$. Similarly, the coordinate transformation $u \mapsto \e^u$, $v \mapsto - \e^v$ results in a class II(a)(ii) spacetime metric with coordinates that cover only the region $r>2M$. In both cases the event horizon, including the fixed point of the KV, is not covered in accordance with the comment made in section \ref{sec:2spaces}.

\subsection{Example: LRS spacetime}
The LRS spacetime (see \cite{tupper1996})
\[
ds^2 = -dt^2 + t^{2a} dr^2 + t^2 d\Omega^2
\]
can be transformed using $t = [(1-a) \tau]^{1/(1-a)}$ into
\[
ds^2 = \tau^{2/(1-a)} [ (1-a)^{-2} \tau^{-2}(-d \tau^2 + dr^2) + d\Omega^2]
\]
and using $\tau = u+v$, $r = v-u$ this takes the form of a class I(a)(ii) spacetime.

\subsection{Example: Moopanar and Maharaj}
In \cite{moopanar2010} an example of a perfect fluid spacetime admitting a ${\cal C}_4 \supset {\cal G}_3$ is given. The metric is
\[
ds^2 = - \case{1}{4} r^2 dt^2 + (\epsilon + c r^2)^{-1} dr^2
+ r^2 [\case{1}{2} \epsilon + h(t)] d\Omega^2
\]
where
\ba
h(t) = A \sin t + B \cos t \:\:\: \mbox{if} \:\:\: \epsilon = -1
\nonumber\\
h(t) = - \case{1}{4} t^2 + At + B \:\:\: \mbox{if} \:\:\: \epsilon = 0
\nonumber\\
h(t) = A \e^t + B \e^{-t} \:\:\: \mbox{if} \:\:\: \epsilon = +1
\nonumber
\ea
with $A$, $B$, $c$, $\epsilon$ being constants. It is easily calculated that the $(V_1,h_1)$ corresponding to $ds^2$ with $h(t)$ as above is not a space of constant curvature and so can admit at most one KV. In \cite{moopanar2010} it is shown that this spacetime admits one proper CKV. This can be shown easily by the method described here. In fact, consider the more general metric of the form
\[
ds^2 = -a^2 r^2 dt^2 + g^2(r) dr^2 + r^2 f^2(t) d\Omega^2
\]
where $a$ is an arbitrary constant and $f(t)$, $g(r)$ are arbitrary functions of their arguments. The 2-space $(V_1,h_1)$ has metric
\[
d\sigma_1^2 = -a^2 f^{-2}(t) dt^2 + r^{-2} g^2(r) f^{-2}(t) dr^2.
\]
Defining $d \tau = af^{-1}(t)dt$ and $d \rho = r^{-1} g(r)dr$ this becomes
\[
d\sigma_1^2 = -d \tau^2 + F^2(\tau) d \rho^2.
\]
This 2-space admits only the KV $\omega = \partial_\rho = r g^{-1}(r) \partial_r$ and reintroducing the multiplying factor $r^2 f^2(t)$ changes this KV into a proper CKV with conformal scalar $\psi = g^{-1}(r)$. Using a further coordinate transformation it is straightforward to show that the spacetime $(M,g)$ is a class II(a)(iv) spacetime. (The II(a)(iv) metric can be transformed into the II(b)(iv) metric {\it locally}, specifically, excluding the spacelike 2-space $u = v = 0$ of the II(b)(iv) metric which corresponds to the fixed point associated with the CKV $\tau$.)

\subsection{Example: Shabbir et al, case 1}
This is case 1 of \cite{shabbir2009} in which the authors find a spacetime admitting proper CKV by direct integration of the CKV equations (\ref{eq:ckv}) and, in the authors' words, some tedious and lengthy calculations. With a slight change of notation, the metric they find is
\[
ds^2 = -r^2 \e^a dt^2 + \e^{bt + c} dr^2 + r^2 d\Omega^2
\]
where $a$, $b \ne 0$ and $c$ are arbitrary constants, so that $d\sigma_1^2 = - \e^a dt^2 + \e^{bt + c} r^{-2} dr^2$. This is a 2-space of constant curvature and so admits a ${\cal G}_3$
\ba
\xi_1 = -2b^{-1} \ln r \partial_t + [\case{1}{2} r (\ln r)^2 + 2 b^{-2} r \e^{a - bt - c} ] \partial_r
\nonumber\\
\xi_2 = -2b^{-1} \partial_t + r \ln r \partial_r,
\qquad
\xi_3 = 2 r \partial_r.
\nonumber
\ea
The corresponding conformal scalars for $(M,g)$ are
\[
\psi_1 = \case{1}{2} \ln r + 2 b^{-2} \e^{a - bt - c},
\qquad
\psi_2 = r \ln r,
\qquad
\psi_3 = 2
\]
so the spacetime admits a ${\cal C}_6 \supset {\cal H}_4 \supset {\cal G}_3$, i.e., the spacetime admits two independent proper CKVs and an HKV. In \cite{shabbir2009}, the authors claim to have found three proper CKV, but did not find $\xi_1$ and missed the fact that $\xi_3$ is a HKV. Using the coordinate transformation
\[
u = \case{1}{2} [ \ln r - 2b^{-1} \e^{a - bt - c}],
\qquad
v = - \case{1}{2} [ \ln r + 2b^{-1} \e^{a - bt - c}]
\]
we find that this is a class I(a)(vi) spacetime.

\subsection{Example: Shabbir et al, case 2}
This example is case 2 of \cite{shabbir2009} which the authors claim admits a ${\cal C}_5 \supset {\cal G}_4$ including the $so(3)$ Lie algebra (\ref{eq:g3sphericalsymmetry}). This is clearly incorrect since $(V_1,h_1)$ cannot admit a maximal ${\cal G}_2$. The line element of $(M,g)$ is
\[
ds^2 = -r^2 dt^2 + \exp N dr^2 + r^2d \Omega^2
\]
where $N = N(r)$ so that the metric $h_1$ is
\be
d \sigma_1^2 = - dt^2 + r^{-2} \exp N dr^2.
\label{eq:shabbir2}
\ee
The transformation
\be
\int r^{-1} \exp (N/2) dr = u + v$, \qquad $t = u - v
\label{eq:shabbir2transformation}
\ee
converts (\ref{eq:shabbir2}) into the flat metric $4 du dv$ with ${\cal G}_3$ basis $\{ \eta_1, \eta_2, \eta_3 \}$. Inverting the transformation we find that the ${\cal C}_6 \supset {\cal G}_4$ of $(M,g)$ consists of (\ref{eq:g3sphericalsymmetry}) and
\ba
\bar \eta_1 = \partial_t, \psi_1 = 0, \qquad \bar \eta_2 = r \exp (-N/2) \partial_r, \psi_2 = \exp (N/2)
\nonumber\\
\bar \eta_3 = \biggl[ \int r^{-1} \exp (N/2) dr \biggr] \partial_t + t r \exp (-N/2) \partial_r,
\psi_3 = r \exp (-N/2)
\nonumber
\ea
where $\eta_1 = \bar \eta_1 + \bar \eta_2$, $\eta_2 = \bar \eta_2 - \bar \eta_1$ and $\eta_3 = \bar \eta_3$. Using the transformation (\ref{eq:shabbir2transformation}) it can be shown that the spacetime $(M,g)$ is a class I(b)(iv) spacetime.

\section{Type $O$ spacetimes}\label{sec:typeo}
The line element of Minkowski spacetime  in null coordinates is
\be
ds^2_M = du dv + \case{1}{4}(u+v)^2 d\Omega^2.
\label{minkowskinullcoordinates}
\ee
The type $O$ 2+2 reducible spacetime $(M, \hat{g})$ is
\be
d\Sigma^2 = 4(u+v)^{-2} du dv + d\Omega^2
\label{typeometric}
\ee
and we remark that $d\Sigma^2 = 4 (u+v)^{-2} ds^2_M$. In the following $\alpha_i, \beta_i, \gamma_i$, $\delta_i$ and $\epsilon_i$, $i=1,2,3$, are arbitrary constants. The spacetime (\ref{typeometric}) admits the ${\cal C}_{15} \supset {\cal G}_6$ consisting of (Note that \cite{carot2002} contains some sign errors.)
\ba
\zeta_{\alpha_1} = \case{1}{2} \sin \theta \sin \phi (u^2 \partial_u + v^2 \partial_v)
- uv (u+v)^{-1} \rho_1
\nonumber\\
\zeta_{\alpha_2} = \sin \theta \sin \phi (-u \partial_u + v \partial_v)
- (u-v) (u+v)^{-1} \rho_1
\nonumber\\
\zeta_{\alpha_3} = - \case{1}{2} \sin \theta \sin \phi (\partial_u + \partial_v)
- (u+v)^{-1} \rho_1
\nonumber\\
\zeta_{\beta_1} = \case{1}{2} \sin \theta \cos \phi (u^2 \partial_u + v^2 \partial_v)
- uv (u+v)^{-1} \rho_2
\nonumber\\
\zeta_{\beta_2} = \sin \theta \cos \phi (-u \partial_u + v \partial_v)
- (u-v) (u+v)^{-1} \rho_2
\nonumber\\
\zeta_{\beta_3} = - \case{1}{2} \sin \theta \cos \phi (\partial_u + \partial_v)
- (u+v)^{-1} \rho_2
\nonumber\\
\zeta_{\gamma_1} = \case{1}{2} \cos \theta (u^2 \partial_u + v^2 \partial_v)
+ uv (u+v)^{-1} \rho_3
\nonumber\\
\zeta_{\gamma_2} = \cos \theta (-u \partial_u + v \partial_v)
+ (u-v) (u+v)^{-1} \rho_3
\nonumber\\
\zeta_{\gamma_3} = - \case{1}{2} \cos \theta (\partial_u + \partial_v)
+ (u+v)^{-1} \rho_3
\nonumber\\
\zeta_{\epsilon_1} = u^2 \partial_u - v^2 \partial_v,
\qquad
\zeta_{\epsilon_2} = u \partial_u + v \partial_v,
\qquad
\zeta_{\epsilon_3} = \partial_u - \partial_v
\nonumber
\ea
and the rotation subalgebra $so(3)$ given by (\ref{eq:g3sphericalsymmetry}), where
\ba
\rho_1 = \cos \theta \sin \phi \partial_\theta + \csc \theta \cos \phi \partial_\phi,
\qquad
\rho_2 = \cos \theta \cos \phi \partial_\theta - \csc \theta \sin \phi \partial_\phi
\nonumber\\
\rho_3 = \sin \theta \partial_\theta.
\nonumber
\ea
Thus, the general CKV is given by
\[
\zeta = \sum_{i=1}^{3} \biggl(\alpha_i \zeta_{\alpha_i} + \beta_i \zeta_{\beta_i}
+ \gamma_i \zeta_{\gamma_i} +  \epsilon_i \zeta_{\epsilon_i} + \delta_i \zeta_{\delta_i} \biggr).
\]
The corresponding conformal scalar is given by
\ba
\fl \hat \psi = (u+v)^{-1} [(\alpha_1
uv+\alpha_2(u-v)+\alpha_3)\sin \theta \sin\phi + (\beta_1
uv+\beta_2(u-v)+\beta_3)\sin \theta \cos\phi
\nonumber\\
+ (\gamma_1 uv+\gamma_2(u-v)+\gamma_3)\cos
\theta ].
\nonumber
\ea
Thus there are five sets of three CKVs: $\zeta_{\alpha_i}, \zeta_{\beta_i}, \zeta_{\gamma_i}$ are nine proper CKVs while $\zeta_{\epsilon_i}$ and $\zeta_{\delta_i}$ are KVs. The conformal scalars corresponding to $(M,g)$ and $(M, \hat{g})$ are related by (\ref{eq:conformalscalarrelationship}), i.e.,
\be
\psi = Z^{-1} (\zeta^u Z_{,u} + \zeta^v Z_{,v}) + \hat \psi.
\label{eq:conformalscalarrelationshiptypeO}
\ee
For a particular value $i = k$, the $u$, $v$ dependence of the CKV $\zeta$ and the conformal scalar $\hat \psi$ are identical for each of $\alpha_k$, $\beta_k$ and $\gamma_k$ and it follows that the condition $\psi = 0$ which gives rise to a KV $\zeta_{\alpha_i}$ will in fact lead to a {\it triplet} of independent KVs $\{ \zeta_{\alpha_i}, \zeta_{\beta_i}, \zeta_{\gamma_i} \}$. Note that there can be no proper HKVs in the spacetime $(M, g)$ (i.e., $\psi = constant \ne 0$) arising from the $\alpha_i, \beta_i, \gamma_i$ since $Z$ is a function of $u$ and $v$ only and the terms in $\hat \psi$ corresponding to $\alpha_i, \beta_i, \gamma_i$ contain trigonometric functions of $\theta$ and $\phi$. However, HKV {\it can} arise from the $\epsilon_i$. Further, for similar reasons, there can be no KVs arising from linear combinations of CKVs from the {\it different} sets $\alpha_i, \beta_j, \gamma_k$, $i \ne j \ne k$. The spacetime $(M,g)$ will admit a ${\cal C}_{15} \supset {\cal G}_3$ for an arbitrary function $Z(u,v)$.

The basis for the Lie algebra ${\cal C}_{15}$ above is the basis (24) in \cite{keane2000},
up to some constant factors, using a standard coordinate transformation. Specifically,
\ba
\zeta_{\alpha_1} &=& \case{1}{2} K_y, \qquad
\zeta_{\alpha_2} = M_{ty}, \qquad
\zeta_{\alpha_3} = - \case{1}{2} T_y
\nonumber\\
\zeta_{\beta_1} &=& \case{1}{2} K_x, \qquad
\zeta_{\beta_2} = M_{tx}, \qquad
\zeta_{\beta_3} = - \case{1}{2} T_x
\nonumber\\
\zeta_{\gamma_1} &=& \case{1}{2} K_z, \qquad
\zeta_{\gamma_2} = M_{tz}, \qquad
\zeta_{\gamma_3} = - \case{1}{2} T_z
\nonumber\\
\zeta_{\epsilon_1} &=& -K_t, \qquad
\zeta_{\epsilon_2} = -D, \qquad
\zeta_{\epsilon_3} = T_t
\nonumber\\
\zeta_{\delta_1} &=& M_{zx}, \qquad
\zeta_{\delta_2} = M_{yz}, \qquad
\zeta_{\delta_3} = M_{xy}.
\nonumber
\ea
So the 15-dimensional algebra can be put in the form of the standard $so(4,2)$ algebra (25) in \cite{keane2000} as we would expect.

We now perform a similar analysis to that in the type $D$ case. We emphasize that here, in contrast to the type $D$ case, we list both subgroups arising from
\begin{enumerate}

\item[I] CKVs $\zeta$ acting {\it only} on the 2-space $(V_1,h_1)$.

\item[II] CKVs $\zeta$ acting on $(M, \hat{g})$.

\end{enumerate}
Thus class I arises from the properties of the CKVs $\zeta_{\epsilon_i}$ and class II from the properties of the CKVs $\zeta_{\alpha_i}$, $\zeta_{\beta_i}$ and $\zeta_{\gamma_i}$. For each of the CKVs $\zeta$ conditions are given for the function $Z(u,v)$ to give ${\cal G}$ or ${\cal H}$ subalgebras. As remarked above, the situation is identical for $\zeta_{\alpha_i}$, $\zeta_{\beta_i}$ and $\zeta_{\gamma_i}$ so only the conditions for $\zeta_{\alpha_i}$ are listed. Further, the analysis for class I is identical to that for class I(a) carried out in section \ref{sec:2spaces} since the metrics $d\sigma_1^2$ have the same functional form, and the $\zeta_{\epsilon_i}$ are identical to the corresponding $\xi_i$, having $\hat \psi = 0$. It will not be necessary to list the corresponding conditions again. Similarly we do not list the $so(3)$ subalgebra since it is common to all.

The following shorthand notation will be used:
\ba
U_\pm = (u + v)(a^2 \pm uv)^{-1}, \qquad V_\pm = (u - v)(a^2 \pm uv)^{-1}
\nonumber\\
W_\pm = (u^2 \pm a^2)^{-1} (v^2 \pm a^2)^{-1}.
\label{eq:shorthandnotation}
\ea
The quantity $a$ is an arbitrary non-zero constant and $F$ denotes an arbitrary function of its arguments. The analysis corresponding to class II gives the following.

\begin{enumerate}

\item $\zeta_{\alpha_3} \in {\cal G}(V_1)$ if $Z = (u + v) F(u - v)$.

\item $\zeta_{\alpha_1} + a^2 \zeta_{\alpha_3} \in {\cal G}(V_1)$ if $Z = (u + v) W^{1/2}_- F(V_-)$.

\item $\zeta_{\alpha_1} - a^2 \zeta_{\alpha_3} \in {\cal G}(V_1)$ if $Z = (u + v)W^{1/2}_+ F(V_+)$.

\end{enumerate}
Again, those classes not listed are equivalent to one of the above via an appropriate coordinate transformation.

The preceding analysis for classes I and II takes no account of the fact that class I (or a subclass)
may admit further symmetry either {\it automatically} or in {\it special cases}. Any further symmetry must correspond to a subclass of class II and since the resulting class arises {\it simultaneously} in both I and II, it shall be designated class III. Class III can be determined by considering the common functional forms $Z$ in both I and II, or by inserting the functions $Z$ directly into (\ref{eq:conformalscalarrelationshiptypeO}). Following the latter method, the pertinent CKV components and the conformal scalar $\hat \psi$ are
\ba
\fl \zeta^u = \case{1}{2} [P(\alpha) \sin \theta \sin \phi + P(\beta) \sin \theta \cos \phi
+ P(\gamma) \cos \theta] - \epsilon_1 u^2 + \epsilon_2 u - \epsilon_3
\nonumber\\
\fl \zeta^v = \case{1}{2} [Q(\alpha) \sin \theta \sin \phi + Q(\beta) \sin \theta \cos \phi
+ Q(\gamma) \cos \theta] + \epsilon_1 v^2 + \epsilon_2 v + \epsilon_3
\nonumber\\
\fl \hat \psi = (u+v)^{-1} [R(\alpha) \sin \theta \sin \phi + R(\beta) \sin \theta \cos \phi
+ R(\gamma) \cos \theta]
\nonumber
\ea
where $P(\alpha) = \alpha_1 u^2 - 2 \alpha_2 u - \alpha_3$, $Q(\alpha) = \alpha_1 v^2 + 2 \alpha_2 v - \alpha_3$, $R(\alpha) = \alpha_1 uv + \alpha_2 (u - v) + \alpha_3$ etc. Writing
$L = \case{1}{2} \ln Z$, equation (\ref{eq:conformalscalarrelationshiptypeO}) is
\ba
\psi &= [ P(\alpha) L_{,u} + Q(\alpha) L_{,v} + (u+v)^{-1} R(\alpha) ]
\sin \theta \sin \phi
\nonumber\\
&+ [ P(\beta) L_{,u} + Q(\beta) L_{,v} + (u+v)^{-1} R(\beta) ]
\sin \theta \cos \phi
\nonumber\\
&+ [ P(\gamma) L_{,u} + Q(\gamma) L_{,v} + (u+v)^{-1} R(\gamma) ] \cos \theta
\nonumber\\
&+ 2 (\epsilon_1 u^2 + \epsilon_2 u + \epsilon_3) L_{,u}
+ 2 (- \epsilon_1 v^2 + \epsilon_2 v - \epsilon_3) L_{,v}.
\nonumber
\ea
So, for an HKV, proper or not, we must have the first three brackets equal to zero and the last bracket equal to a constant $\psi_0$, i.e.,
\ba
P(\alpha) L_{,u}  + Q(\alpha) L_{,v} + (u+v)^{-1} R(\alpha) = 0
\label{eq:overlaps1}
\\
(\epsilon_1 u^2 + \epsilon_2 u + \epsilon_3) L_{,u}
+ (- \epsilon_1 v^2 + \epsilon_2 v - \epsilon_3) L_{,v} = \case{1}{2} \psi_0
\label{eq:overlaps2}
\ea
the equations for $\beta$ and $\gamma$ being identical to those for $\alpha$. It is necessary to solve equations (\ref{eq:overlaps1}) and (\ref{eq:overlaps2}) for classes I and II. This is an extremely laborious task and only the final results are presented.

\begin{table}
\caption{\label{tab:typeO}The type $O$ metrics and corresponding algebras. The function $F$ is an {\it arbitrary} function of its arguments. The $so(3)$ isometry subalgebra (\ref{eq:g3sphericalsymmetry}) is not listed since it is common to all classes. For a spherically symmetric spacetime admitting a homothety, the homothety algebra contains an isometry subalgebra with dimension one less than the dimension of the homothety algebra, i.e., ${\cal H}_r \supset {\cal G}_{r-1}$. The notation $\zeta \in {\cal H}$ denotes that $\zeta$ is a proper HKV with $\psi = c \ne 0$, and $\zeta \in {\cal S}$ denotes that $\zeta$ is a proper SCKV. The quantities $U_\pm, V_\pm, W_\pm$ are defined in equations (\ref{eq:shorthandnotation}).
}
\footnotesize\rm
\begin{tabular*}{\textwidth}{@{}l*{15}{@{\extracolsep{0pt plus12pt}}l}}
\br
Class & & Algebra & & Metric $ds^2$ \\
\mr

I & (i) & ${\cal C}_{15} \supset {\cal G}_6$ & $\zeta_{\epsilon_1}, \zeta_{\epsilon_2}, \zeta_{\epsilon_3} \in {\cal G}$ & $d\Sigma^2 = 4(u+v)^{-2} du dv + d\Omega^2$ \\

& (ii) & ${\cal C}_{15} \supset {\cal G}_4$ & $\zeta_{\epsilon_3} \in {\cal G}$ & $F^2(u+v) d\Sigma^2$ \\

& (iii) & ${\cal C}_{15} \supset {\cal H}_5$ & $\zeta_{\epsilon_2} \in {\cal H}$, $\zeta_{\epsilon_3} \in {\cal G}$ & $(u+v)^c d\Sigma^2$ \\

& (iv) & ${\cal C}_{15} \supset {\cal H}_4$ & $\zeta_{\epsilon_3} \in {\cal H}$ & $\e^{-2cv} F^2(u+v) d\Sigma^2$ \\

& (v) & ${\cal C}_{15} \supset {\cal G}_4$ & $\zeta_{\epsilon_1} + a^2 \zeta_{\epsilon_3} \in {\cal G}$ & $F^2(U_-) d\Sigma^2$ \\

& (vi) & ${\cal C}_{15} \supset {\cal H}_4$ & $\zeta_{\epsilon_1} + a^2 \zeta_{\epsilon_3} \in {\cal H}$ & $\exp[2ca^{-1} \arctan(v/a)] F^2(U_-) d\Sigma^2$ \\

& (vii) & ${\cal C}_{15} \supset {\cal G}_4$ & $\zeta_{\epsilon_1} - a^2 \zeta_{\epsilon_3} \in {\cal G}$ & $F^2(U_+) d\Sigma^2$ \\

& (viii) & ${\cal C}_{15} \supset {\cal H}_4$ & $\zeta_{\epsilon_1} - a^2 \zeta_{\epsilon_3} \in {\cal H}$ & $[(a+v)/(a-v)]^{-c/a} F^2(U_+) d\Sigma^2$ \\

& (ix) & ${\cal C}_{15} \supset {\cal G}_3$ & - & $F^2(u,v) d\Sigma^2$ \\

& & & & \\

II & (i) & ${\cal C}_{15} \supset {\cal G}_6$ & $\zeta_{\alpha_3},\zeta_{\beta_3},\zeta_{\gamma_3} \in {\cal G}$ & $(u+v)^2 F^2(u-v) d\Sigma^2$ \\

& (ii) & ${\cal C}_{15} \supset {\cal G}_6$ & $\zeta_{\alpha_1} + a^2 \zeta_{\alpha_3}, \zeta_{\beta_1} + a^2 \zeta_{\beta_3},$ & $(u+v)^2 W_- F^2(V_-) d\Sigma^2$ \\

& & & $\zeta_{\gamma_1} + a^2 \zeta_{\gamma_3} \in {\cal G}$ & \\

& (iii) & ${\cal C}_{15} \supset {\cal G}_6$ & $\zeta_{\alpha_1} - a^2 \zeta_{\alpha_3},\zeta_{\beta_1} - a^2 \zeta_{\beta_3},$ & $(u+v)^2 W_+ F^2(V_+) d\Sigma^2$ \\

& & & $\zeta_{\gamma_1} - a^2 \zeta_{\gamma_3} \in {\cal G}$ & \\

& & & & \\

III & (i) & ${\cal C}_{15} \supset {\cal G}_{7}$ & $\zeta_{\alpha_2}, \zeta_{\beta_2}, \zeta_{\gamma_2}, \zeta_{\epsilon_2} \in {\cal G}$ & $(u+v)^2 (uv)^{-1} d\Sigma^2$ \\

& (ii) & ${\cal C}_{15} \supset {\cal G}_{7}$ & $\zeta_{\alpha_1} - a^2 \zeta_{\alpha_3},\zeta_{\beta_1} - a^2 \zeta_{\beta_3},$ & $(u+v)^2 W_+ d\Sigma^2$ \\

& & & $\zeta_{\gamma_1} - a^2 \zeta_{\gamma_3},
\zeta_{\epsilon_1} + a^2 \zeta_{\epsilon_3} \in {\cal G}$ & \\

& (iii) & ${\cal C}_{15} \supset {\cal H}_{7}$ & $\zeta_{\epsilon_2} \in {\cal H}, \zeta_{\alpha_2}, \zeta_{\beta_2}, \zeta_{\gamma_2} \in {\cal G}$ & $(u+v)^2 (uv)^{c-1} d\Sigma^2$ \\

& (iv) & ${\cal C}_{15} \supset {\cal H}_{7}$ & $\zeta_{\epsilon_2} \in {\cal H}, \zeta_{\alpha_3}, \zeta_{\beta_3}, \zeta_{\gamma_3} \in {\cal G}$ & $(u+v)^2 (u-v)^{2(c-1)} d\Sigma^2$ \\

& (v) & ${\cal C}_{15} \supset {\cal H}_{7}$ & $\zeta_{\epsilon_3} \in {\cal H}, \zeta_{\alpha_3}, \zeta_{\beta_3}, \zeta_{\gamma_3} \in {\cal G}$ & $(u+v)^2 \e^{c(u-v)} d\Sigma^2$ \\

& (vi) & ${\cal C}_{15} \supset {\cal H}_7$ & $\zeta_{\alpha_1} - a^2 \zeta_{\alpha_3},\zeta_{\beta_1} - a^2 \zeta_{\beta_3},$ & $(u+v)^2 W_+ \exp[ca^{-1} \arctan (a V_+ )] d\Sigma^2$ \\

& & & $\zeta_{\gamma_1} - a^2 \zeta_{\gamma_3} \in {\cal G}$, & \\

& & & $\zeta_{\epsilon_1} + a^2 \zeta_{\epsilon_3} \in {\cal H}$ & \\

& (vii) & ${\cal C}_{15} \supset {\cal G}_{10}$ & $\zeta_{\alpha_2}, \zeta_{\beta_2}, \zeta_{\gamma_2}, \zeta_{\alpha_1} - a^2 \zeta_{\alpha_3},$ & $U^2_+ d\Sigma^2$ \\

& & & $\zeta_{\beta_1} - a^2 \zeta_{\beta_3}, \zeta_{\gamma_1} - a^2 \zeta_{\gamma_3},$ & \\

& & & $\zeta_{\epsilon_1} - a^2 \zeta_{\epsilon_3} \in {\cal G}$ & \\

& (viii) & ${\cal C}_{15} \supset {\cal G}_{10}$ & as above with $a^2 \mapsto -a^2$ & $U^2_- d\Sigma^2$ \\

& (ix) & ${\cal S}_{15} \supset {\cal H}_{11}$ & $\zeta_{\alpha_2}, \zeta_{\alpha_3}, \zeta_{\beta_2}, \zeta_{\beta_3},\zeta_{\gamma_2}, \zeta_{\gamma_3},$& $\case{1}{4} (u+v)^2 d\Sigma^2$ \\

& & & $\zeta_{\epsilon_3} \in {\cal G},\zeta_{\epsilon_2} \in {\cal H},$ & \\

& & & $\zeta_{\alpha_1}, \zeta_{\beta_1}, \zeta_{\gamma_1}, \zeta_{\epsilon_1} \in {\cal S}$ & \\

\br
\end{tabular*}
\end{table}

The complete classification for the type $O$ spacetimes $(M,g)$ is given in table \ref{tab:typeO}. The class I(ix) spacetime
\[
ds^2 = F^2(u,v) d\Sigma^2
\]
is the {\it general} spherically symmetric type $O$ spacetime, for an arbitrary function $F$, and admits no further isometries or homotheties. The spherically symmetric Stephani universes \cite{barnes1998} are the spherically symmetric type $O$ expanding perfect fluid spacetimes and are a special case of this general class.

\subsection{Example: Bertotti-Robinson Electromagnetic Universe}
The Bertotti-Robinson electromagnetic universe \cite{bertotti1959, robinson1959} is the unique conformally flat source-free solution of the non-null Einstein-Maxwell equations. One form of the metric is
\[
ds^2 = r^{-2}(-dt^2 + dr^2 + r^2d\Omega^2).
\]
Putting $r = u + v$ and $t = u - v$, this becomes precisely the metric (\ref{typeometric}). Thus we know immediately that this spacetime is the class I(i) and admits the ${\cal C}_{15} \supset {\cal G}_6$ given above.

\subsection{Example: Robertson-Walker spacetimes}
The general Robertson-Walker spacetimes are given by classes II(i), II(ii) and II(iii) for $k=0$, $k < 0$ and $k >0$ respectively, as is shown presently: Under the coordinate transformation
\be
u = (r + t)/2, \qquad v = (r - t)/2
\label{eq:transformationuvrt}
\ee
the metric of class II(i) is
\[
ds^2 = R^2(t) [-dt^2 + dr^2 + r^2 d\Omega^2]
\]
which is clearly the general $k = 0$ Robertson-Walker spacetime; Following the procedure in section 3 of \cite{keane2000}, the Robertson-Walker spacetimes for arbitrary $k$ can be written in the manifestly conformally flat form
\ba
ds^2 = S^2(\tau) G^2(\tau', r', k)[-d\tau'^2 + dr'^2 + r'^2 d\Omega^2]
\nonumber\\
G^2(\tau', r', k) = {( {[(1+k\tau'^2 / 2) + (1+kr'^2 /2)]^2} - k^2 \tau'^2 r'^2 )^{-1}}
\nonumber
\ea
where $S$ is an arbitrary function of $\tau$ defined through equations (17) and (18) of \cite{keane2000}. Using the coordinate transformation $\tau' = u - v$, $r' = u + v$ and $k = a^{-2}$ it follows that $\tau = a \arctan(a V_+)$ and the metric is precisely that of class II(iii). The same procedure with $k = -a^{-2}$ transforms the general $k<0$ Robertson-Walker spacetime into the class II(ii) metric. The class II metrics represent the general Robertson-Walker spacetimes in manifestly conformally flat null coordinates. In analogy with \cite{keane2000} we can express the entire class in terms of a single metric and the continuous spatial curvature parameter $k$. Using metric II(iii) as the template, under the inversion transformation $u \mapsto u^{-1}$, $v \mapsto v^{-1}$
\be
ds^2 = 4 (1 + ku^2)^{-1} (1 + kv^2)^{-1} F^2[(u-v)/(1 + kuv)] ds^2_M
\label{eq:rwcfk}
\ee
where $ds^2_M$ is Minkowski spacetime (\ref{minkowskinullcoordinates}). This metric was previously discovered by Sopuerta \cite{sopuerta1998}. This expression applies for arbitrary real values of $k$ and arbitrary functions $F$ and is an alternative to those manifestly conformally flat Robertson-Walker metrics given in \cite{keane2000}, \cite{tauber1967} - \cite{ibison2007}.

It is of interest that several alternative forms for the Robertson-Walker metric arise in the analysis. (They have been eliminated from the classification since they correspond to existing classes expressed in different coordinate systems.) In particular the metric
\be
ds^2 = (u+v)^2 F^2(uv) d\Sigma^2
\label{eq:RobertsonWalkerknegalternative}
\ee
is the general $k<0$ Robertson-Walker spacetime. This can be transformed directly into the metric II(ii) using the coordinate transformation
\[
u \mapsto (a - u)/(a + u), \qquad v \mapsto (a + v)/(v - a).
\]
However, it is more illustrative to use the coordinate transformation (\ref{eq:transformationuvrt}) which transforms the metric (\ref{eq:RobertsonWalkerknegalternative}) into
\be
ds^2 = G^2(t^2 - r^2) [-dt^2 + dr^2 + r^2 d\Omega^2]
\label{eq:tauberkneg}
\ee
where $G^2(t^2 - r^2) = F^2(uv)$. Using the coordinate transformation (1.7d) and the relation (1.6d) of Tauber \cite{tauber1967} this can be cast into the standard form
\[
ds^2 = -dt^2 + R^2(t) \biggl(1 + \frac{k}{4} r^2 \biggr)^{-2}(dr^2 + r^2 d\Omega^2),
\qquad k <0.
\]

Classes III(i) and III(ii) are the Einstein static spacetimes for $k<0$ and $k >0$ respectively. In particular, under the coordinate transformation (\ref{eq:transformationuvrt}), class III(i) is
\[
ds^2 = (t^2 - r^2)^{-1} [-dt^2 + dr^2 + r^2 d\Omega^2]
\]
which is obviously a special case of (\ref{eq:tauberkneg}). Those Robertson-Walker spacetimes admitting an ${\cal H}_7$ are classes III(iii), III(vi) for $k<0$ and $k >0$ respectively, and both III(iv) and III(v) for $k = 0$. (These two $k = 0$ metrics correspond to distinct spacetimes, having distinct ${\cal H}_7$ Lie algebras.) The corresponding conditions on the Robertson-Walker scale factor can be found in tables 1 and 2 of \cite{maartens1986}. Classes III(vii) and III(viii) are the de Sitter and anti de Sitter spacetimes respectively and class III(ix) is Minkowski spacetime. Using the coordinate transformation (\ref{eq:transformationuvrt}) Minkowski spacetime is $ds^2_M = -dt^2 + dr^2 + r^2 d\Omega^2$ and classes III(vii) and III(viii) take on the familiar forms $[\case{1}{4} (r^2 - t^2) + a^2]^{-2}ds^2_M$ and $[\case{1}{4} (r^2 - t^2) - a^2]^{-2}ds^2_M$ respectively.

\section{Conclusion}
A complete classification of locally spherically symmetric Lorentzian spacetimes is given in terms of their local conformal symmetries. This generalizes previous work on isometries and on conformal symmetries in static spacetimes. Previous work in this area has been based upon direct integration of the conformal Killing equations involving, in general, many functions of two independent variables. By taking advantage of the work on 2+2 reducible spacetimes the classification problem is reduced to the various functional forms of a {\it single} function of two independent variables and an underlying 2+2 reducible metric. Further, the classification is based on types $D$ or $O$ separately from the outset in contrast to most existing schemes. Thus, the {\it general} solution of the classification problem has been given in terms of canonical metric types and the associated conformal Lie algebras. The classification is independent of the field equations.

Several alternative forms for the Robertson-Walker metric are given in terms of null coordinates and of particular interest is the manifestly conformally flat metric (\ref{eq:rwcfk}) which applies for {\it arbitrary} real values of spatial curvature.

\section*{Acknowledgments}
We would like to thank Graham Hall for comments on the first draft of this paper
and we would like to thank the referees for their suggestions.

\end{document}